\documentclass[12pt]{article}
\usepackage{amsmath,amssymb,pgf,pgfarrows,pgfnodes,subfig,float,appendix, hyperref}
\usepackage[margin=0.7in]{geometry}

\title{{\rm\footnotesize \qquad \qquad \qquad \qquad \qquad \ \qquad \qquad \qquad \ \ \ \ \ \                    RUNHETC-2013-1     
SCIPP 13/10}\vskip.5in     Lectures on Holographic Space-time}
\author{Tom Banks\\
Department of Physics and SCIPP\\
University of California, Santa Cruz, CA 95064\\
{\it and}\\
Department of Physics and NHETC\\
Rutgers University, Piscataway, NJ 08854\\
E-mail: \href{mailto:banks@scipp.ucsc.edu}{banks@scipp.ucsc.edu}}
\begin{document}
\maketitle

\begin{abstract}
Summary of three talks on the Holographic Space-Time (HST) models of early universe cosmology, particle physics, and the asymptotically de Sitter final state of our universe.
\end{abstract}
\section{Introduction}
This article summarizes talks I gave at the Davis Conference on Cosmology, in May of 2013, the Tales of Lambda conference at the University of Nottingham in July of 2013, and the Benasque workshop on String Theory at the Centro de Ciencias Pedro Pascual, also in July of 2013.  I'd like to thank the organizers of all of these conferences for inviting me to attend, and particularly to thank those of the participants of the Benasque workshop who voted to hear my talk.  The ideas described here were mostly developed in collaboration with W. Fischler.  Only the connections between holography, the spinor bundle on the holographic screen, and supersymmetry, for whatever they are worth, were invented by me.
I deliberately omit the HST discussion of the Firewall paradox in black hole physics, which will be re-examined in a forthcoming paper\cite{fw3}.

For more than a decade, Willy Fischler and I have been working on a formalism called Holographic Space-time (HST), whose intent was to provide a quasi-local theory of quantum gravity, which generalizes string theory beyond asymptotically flat and AdS backgrounds.  HST has provided us with a complete and non-singular model of cosmology, going from the Big Bang to the end of the inflationary era.  This model resolves all of the issues alluded to by Frederik Denef in his talk at Benasque, as well as the trans-Planckian and initial condition problems of conventional inflation.  Its predictions for current data cannot be distinguished from those of inflationary models, but careful measurements of tensor fluctuations can differentiate between them.
HST also provides a theory of stable dS space, which explains what will apparently be the ultimate fate of our universe.  

What is missing so far is a complete theory of particle physics, which is necessary for the description of the universe from the end of inflation until the present day.  In my Benasque talk, I outlined recent progress towards such a theory.  

The basic geometrical object, for which HST provides a quantum avatar, is a causal diamond.  The boundary of a causal diamond has two smooth metrics, describing its past and future light cones.  The two metrics agree on the intersection of the two light cones. They have the form
$$ds^2 = - g^{\pm}_{ij} (u) dx^i dx^j ,$$ where $u$ is a null coordinate and $g_{ij}$ is an Euclidean $d - 2$ metric. The holographic screen of a diamond is the maximal area, space-like $d-2$ surface on the boundary of the diamond.  The strong form of the covariant entropy bound says that this area, in Planck units, is four times the log of the dimension of the Hilbert space, which describes all experiments that can be performed in that diamond.

A time-like trajectory can be viewed as a nested sequence of causal diamonds. A smaller diamond is a tensor factor in the larger diamond's Hilbert space.  Causality is imposed on the Hamiltonian, which propagates the system in proper time along the trajectory, by insisting that the part of the dynamics corresponding to the segment of the trajectory in a smaller diamond factorizes according to this tensor factorization.  This means that the Hamiltonian describing what goes on inside the causal diamond accessible at any given time, is necessarily time dependent, including more and more degrees of freedom (DOF) as it explores a larger segment of the trajectory's proper time.  At discrete intervals of time, the Hilbert space on which the proper part of the Hamiltonian acts, gets larger.  In geometries with rotation invariant dynamics, the discrete time intervals are of Planck size. In the limit of infinite proper time, (in one or both directions) one gets to a Hilbert space of maximum dimension.  The asymptotics of proper time/(dimension = exponential of area) is determined by the value of the cosmological constant (c.c.), remaining finite for positive c.c. and going to infinity at a finite proper time determined by the negative c.c. .   The case of vanishing c.c. is intermediate between these two.  The area of the holographic screen for vanishing c.c. increases like $t^{d-2}$, asymptotically in proper time.

A full quantum space-time is determined by an infinite number of quantum systems, each representing a different time-like trajectory.  A minimal set of trajectories is parametrized by a topological lattice, which determines the topology of the non-compact initial value surface. For nearest neighbors, the overlap of causal diamonds of identical area is such that the maximal causal diamond in the overlap has precisely one pixel less area than the individual diamonds.  The notion of pixel depends on the fuzzy geometry of the compact dimensions, and will be discussed below.  For non-nearest neighbors, the overlaps are part of the specification of the dynamics.  The overlaps, the Hamiltonians of individual trajectories, and the initial states in the individual trajectory Hilbert spaces, are mutually constrained by the requirement that at each time and for each pair of trajectories, the two density matrices in the overlap, determined by the dynamics of independent trajectories, are unitarily equivalent.  It may be that we should impose further restrictions on multiple overlaps.  In the simple models of cosmology that have been worked out, these are automatically satisfied.

A striking feature of this formulation of quantum gravity is that space-time geometry is not a fluctuating quantum variable.  Both the causal structure and the conformal factor of the space-time metric are determined by quantum data.  They come respectively from the commutation properties of operator algebras (as in QFT) and the dimensions of Hilbert spaces. Lorentzian geometry is a way of encoding the causal restrictions on the time evolution along individual world-lines, the overlapping information shared by pairs of world lines, and the dimensions of the tensor factors in Hilbert spaces, which are implied by the overlap conditions.  
String theorists may reject this possibility, since they know that string theory contains gravitons.

The essential point of view, which reconciles HST with string theory, is Jacobson's\cite{ted} proposal that Einstein's equations are the TH(ermodynamic) E(ffective) F(ield) T(heory) of a quantum system, which obeys the Bekenstein-Hawking law.  Jacobson argued that if the maximally accelerated Rindler-Unruh observer in a Lorentzian space time saw entropy proportional to area in a succession of small causal diamonds along its trajectory, then the first law of thermodynamics implied Einstein's equations, apart from a determination of the c.c. .   In other words, Einstein's equations are like the equations of hydrodynamics.  They should only be quantized when one is studying small low energy excitations of a system which has a ground state (like string theory in asymptotically flat or AdS space-times) but will be valid as classical equations for the coarse grained behavior of high entropy states of quantum gravity, in regimes where QU(antum) E(ffective) F(ield) T(heory) is not a good approximation.  The success of QUEFT in reproducing properties of real theories of QG in appropriate regimes has been extraordinarily useful, but has also led us to a false sense of security about the use of QUEFT to model more localized physics.   In addition, Jacobson's thermodynamic relation of the classical Einstein equations to any theory of gravity incorporating the Bekenstein-Hawking law, ignoring Jacobson's injunction about the impropriety of quantizing hydrodynamics in situations of high entropy, has increased our tendency to rely on quantum field theoretic arguments in places where they are inappropriate.  The rigorous parts of string theory, those which involve small fluctuations about a ground state that is symmetric under the Poincare or Anti-de Sitter groups, all involve situations in which the entropy is low, and quantization of hydrodynamics gives the right low energy fluctuations.  Other questions, like the String Landscape, and stringy inflation (as well as the usual QUEFT discussion of inflation) are, according to HST, outside the circumscribed region in which QUEFT provides a good approximation to HST.  

\section{Summary of HST}

HST is defined by an infinite set of quantum systems, labeled by the points of a topological lattice with the topology of flat $d-1$ dimensional space.  All of the explicit constructions employ a hypercubic lattice, but it is clear that any regular lattice will serve as well, and probably it is sufficient to label the systems by the zero simplices of a $d-1$ dimensional simplicial complex whose topology is that of flat space.  This complex describes the topology of a Cauchy surface in space-time, and the individual systems are taken to be quantum representations of the proper time dynamics along individual time-like trajectories penetrating this surface.  

For Big Bang space-times, we choose the Cauchy surface to be the Big Bang hypersurface and the past tip of every diamond lies on this surface. The time-like trajectories, which are not generally geodesics, are synchronized so that a fixed instant in proper time corresponds to the same area for the causal diamond back to the Big Bang.  In fact, the actual models we study correspond to flat FRW cosmologies, so that the trajectories are all geodesics.  Each quantum system has a time evolution operator $U(T,0)$. In time symmetric space-times, like Minkowski, AdS and eternal dS, we extend
the trajectories out symmetrically from a single plane of time symmetry, and have a sequence of operators $U(T, - T)$.  

In order to incorporate causal evolution into HST we insist that the Hamiltonian at any time can be decomposed as $H(t) = H_{in} (t) + H_{out} (t)$, where the two terms depend on independent anti-commuting DOF, and are even functions of them. We also insist that as the time interval gets longer, at discrete times $t_n$, the Hamiltonian $H_{in} (t_n)$ couples together more DOF, while $H_{out} (t_n) $ acts on fewer.  This implements the intuitive notion of causality.

We could try to model systems of this type in QUEFT, by choosing time-like trajectories, and foliations of their causal diamonds, and defining $H_{in} (t)$ for a trajectory by integrating $T_{\mu\nu} v^{\mu} v^{\nu}$ over the small region that is added to the diamond between time $t$ and $t + \delta$.  $T_{\mu\nu}$ is the stress tensor, and $v^{\mu}$ is the parallel transport of the tangent to the trajectory to points on the space-like slice contained in the disjoint union $D_2 - D_1 \cap D_2$ of the two causal diamonds.  There are obvious choices involved here, the most serious of which is the choice of a smoothing function of compact support.  The UV divergences of QUEFT restricted to bounded regions, like the infinite entanglement entropy per unit boundary area, make this definition pretty useless.  The holographic principle is the proper UV regulator for these divergences.  It is a cutoff on {\it entropy}.

Below, we will define a {\it pixel Hilbert space}, ${\cal P}$, which represents all of the information about compact spatial dimensions.  The entire Hilbert space of the trajectory is called ${\cal H} (\infty )$.  At time $t_n$, the evolution operator $U(t_n , t_0 )$ (where $t_0 = 0$ for Big Bang, and $- t_n$ for time symmetric, space-times), factorizes as $U_{in} (t_n , t_0) \otimes U_{out} (t_n, t_0) $.  $U_{in} (t_n, t_0)$ operates in a Hilbert space ${\cal H} (t_n)$ and $U_{out} (t_n, t_0)$ in the tensor complement of ${\cal H} (t_n)$ in 
${\cal H} (\infty )$.  For space-times which are small deformations of the maximally symmetric spaces with a given value of the c.c., ${\cal H} (t_n ) = \otimes {\cal P}^{(t_n)^{d - 2} }$ when $1 \ll t_n \ll |\Lambda |^{- \frac{1}{d}}$.  Here, everything is in Planck units and we've assumed $|\Lambda | \ll 1$.
If the c.c. is negative, then ${\cal H} (t_n ) = {\cal H} (\infty ) = \infty$ at a finite value of $t_n$, of order the scale set by the c.c. .  If $\Lambda > 0$, ${\cal H} (\infty )$ is finite dimensional and the fixed Hilbert space is achieved at a finite value of $t_n$ of order the scale set by the c.c. .   
Referring to Jacobson's thermodynamic interpretation of Einstein's equations, the c.c. was not determined by local thermodynamics.  We see that instead, it acts, as a boundary condition relating asymptotics in proper time to asymptotics in area.

We have a quantum system with the above properties for each lattice point ${\bf x}$ on the initial Cauchy surface.  For any pair of trajectories and any proper time $t_n$ we specify an overlap Hilbert space ${\cal O} (t_n , {\bf x,y})$ , which is a tensor factor in both ${\cal H} (t_n , {\bf x})$ and ${\cal H} (t_n , {\bf x})$.  The dynamics and choice of initial state in each system gives us two density matrices $ \rho (t_n , {\bf x} , {\bf y}) $ and $ \rho (t_n , {\bf y} , {\bf x}) $ on ${\cal O}$.  {\it The fundamental dynamical consistency condition is that $$ \rho (t_n , {\bf x} , {\bf y}) =U(t_n , {\bf x,y}) \rho (t_n , {\bf y} , {\bf x}) U(t_n , {\bf x,y})^{\dagger} .$$} For nearest neighbors on the lattice, ${\cal O} (t_n , {\bf x,y}) = {\cal H} (t_n - 1) $ .  Note that, since we've chosen our discretization of time to be a single Planck time, this implies that nearest neighbor trajectories are, at all times, a Planck distance apart\footnote{In principle, we could make the time and space resolution smaller, since we only have to add a two dimensional tensor factor at each time step.  However, all the models we consider have $SO(d-1)$ rotation invariance in $d$ space-time dimensions, and in order to preserve this symmetry we must add a full angular momentum multiplet to the spinor variables discussed below.  This makes the time unit of order the Planck time.  It's certainly possible to invent continuously varying Hamiltonians, which interpolate between our discrete evolution operators, but it is not clear whether there could be observational significance to the difference between two different continuous interpolations.}.  There may be independent consistency conditions on multiple overlaps, but in the examples we've considered, all such conditions are automatically satisfied. 

This infinite collection of consistently comparable quantum systems determines the causal structure and conformal factor of a $d$ dimensional Lorentzian geometry.  The geometry is determined because the quantum mechanics defines a set of causal diamonds via the tensor factorization of Hilbert spaces and the principal that commuting operators are space-like separated.  The Bekenstein-Hawking formula determines the conformal factor.  Note that the space-time metric is not generally a fluctuating quantum variable. This is consistent with Jacobson's view that it encodes the hydrodynamics of the quantum system.  We will see that, in asymptotically flat and AdS spaces, at least when the AdS radius is large, the multiple quantum systems of HST become identified in the limit of large proper time.  The resulting systems have a unique ground state, and gravitons are included among the excitations of that ground state.  As usual, hydrodynamics is only quantized when one is considering low energy excitations around a ground state.

The quantum variables of HST are quantized excitations of the geometry of holoscreens.  A holographic screen of finite area is a compact Euclidean manifold.  Connes has shown that the geometry of such a manifold is encoded in the properties of the Dirac equation\cite{connes}. We will offer a quantum mechanical version of Connes' ideas, which also incorporates ideas from string theory about the relation between wrapped BPS branes and the geometry of compact manifolds.  Our quantum variables also incorporate ideas of Cartan and Penrose about spinors and null directions, and clarify the role of the BMS group in the quantum theory of asymptotically flat space-time.  

In any number of space-time dimensions, the Cartan-Penrose equation, $$\bar{\psi} \gamma_{\mu} \psi (\gamma^{\mu} )_{\alpha}^{\beta} \psi_{\beta} = 0,$$ defines a null direction, $\bar{\psi} \gamma_{\mu} \psi$.  $\psi$ itself is a null plane spinor for that direction, which determines a set of transverse hyperplanes of dimension $2 \ldots d-2$.  Thinking naively about a pixel on a holographic screen, this is exactly the information that encodes its local embedding into space-time.  For the screen as a whole, the totality of this information is a section of the spinor bundle over the screen.  In dimensions $2,3,4,6$ and $10$, this equation is solved by every spinor in an irreducible representation of the Lorentz group, but the connection between what Cartan calls {\it pure} spinors, and null directions is quite general.

We will consider screens whose emergent classical geometry is $S^{d-2} \times K$, where the sphere has size that varies with $t_n$, while the compact manifold $K$ is fixed.  The holographic principle tells us that when the radius of the sphere is finite in Planck units, the dimension of the Hilbert space is finite.  This means that the ``spinor bundle" must have finite dimension, and that the finite space of sections must satisfy quantum commutation relations that are realizable in a finite dimensional positive metric Hilbert space.   

An obvious way to cut the spinor bundle down to size, is to impose an eigenvalue cutoff on the Dirac equation.  The Dirac equation on a compact Euclidean D-fold has a discrete spectrum of pairs of eigenvalues $\pm p_n$ with unbounded range.  For large eigenvalue, the degeneracy is $|p|^D$.  Label the eigen-sections $\psi_n$.  We will quantize them by hypothesizing a superalgebra
 $$ [\psi_m , \psi_n ]_+ = Z_{mn}, $$
 $$ [Z_{mn} , \psi_k ] = f_{mn;k} \psi_k .$$  The $Z_{mn}$ satisfy an ordinary Lie algebra.  We will insist that this algebra has a finite dimensional unitary representation, which is irreducible w.r.t. the $\psi_k$.  All states are generated by acting with the fermionic generators on a single state.   It follows that for large eigenvalue cutoff, the entropy of this system scales like $|p|^D$.  The strong holographic principle says that this should be identified with the volume of the manifold $S^{d-2} \times K$ in $d + {\rm dim} (K) = D +2$ dimensional Planck units.  For a sphere of variable radius and a fixed manifold $K$, the spinor bundle is the product of the bundle over the sphere, and that over $K$.  An eigenvalue cutoff on the sphere's Dirac operator is the same as an angular momentum cutoff.   If we introduce a generalized label $j$ for the $d-2$ dimensional spinor spherical harmonics with fixed value of total angular momentum $\leq N$ and Cartan generators of $SO(d - 1)$,  then we can quantize in a rotationally invariant way by writing
 $$[\psi_{ j} (P) , \psi_{ k} (Q) ]_+ =  \delta_{jk} Z_{PQ} ,$$ where $P$ and $Q$ run over the cutoff spectrum of the Dirac operator on $K$.  

If $K$ has a covariantly constant spinor, this is a zero mode of the Dirac operator.  We postulate that $ Z_{00}$ is a positive c-number, so that the zero mode generators satisfy an ordinary Clifford algebra.   Denote the eigensections of the Dirac operator on the sphere by $\phi_j (\Omega)$ For large $N$ the operator
$$\psi_{ab} (\Omega_0, \Omega ) \equiv \sum \phi^*_{j \ a}(\Omega_0) \psi_{j \ b},$$ approaches a delta function distribution, $\delta (\Omega , \Omega_0 ) \delta_{ab}$,  localized at  the point $\Omega_0$ on the sphere.  By adjusting $Z_{00}$ in the large $N$ limit we get, for each pair of measurable sections of the spinor bundle $f_a (\Omega)$ and $g_a (\Omega)$
$$[ \psi [f , \Omega_0], \psi [g , \Omega_0]  ]_+ = P f_a (\Omega_0 ) g_a (\Omega_0 ) .$$ Here $$ \psi [f , \Omega_0] \equiv \int \psi_a (\Omega_0, \Omega ) f_a (\Omega ) .$$ $P$ is a positive real number, which is defined in the course of taking the limit.

If we want to describe an asymptotically flat space time, then the kinematic variables should become covariant under the Lorentz group $SO(1,d-1)$, which is the conformal group of the $d-1$ sphere. \footnote{This requirement actually follows from the postulates of HST.  In Minkowski space, the particle sub-factors in the Hilbert spaces of causal diamonds associated with arbitrary geodesics, become identical as proper time goes to infinity.  The consistency conditions then imply that the S-matrices computed along two different geodesics are unitarily equivalent. Since the geodesics are related by Lorentz transformations, this implies that the common Hilbert space carries a unitary representation of the Poincare group, which commutes with the S-matrix.} The conformal Killing spinors of the sphere satisfy
$$ D_M^{ab} S_{\alpha\ b} = e_M^A (\Gamma_A )^{ab} S_{\alpha b} ,$$  where $e_M^A$ is the vielbein on the sphere and $\Gamma_A$ are the $d - 2$ dimensional Euclidean Dirac matrices. This is a conformally covariant equation and the spinors $S_{\alpha a}$ transform according to the Dirac spinor representation of $SO(1, d - 1)$.
The operators $$ Q_{\alpha} \equiv \psi [S_{\alpha\ a}, \Omega_0] ,$$ satisfy 
$$ [Q_{\alpha}, Q_{\beta}]_+ = (\gamma_0 \gamma_{\mu})_{\alpha\beta} P (1, \pm\Omega_0) .$$  The ambiguity in the $\pm$ sign comes from the fact that the commutation relations and the CKS equation are separately invariant under parity (orientation reversal on the sphere).  So we can construct two versions of the SUSY generators, called $Q_{\alpha}^{\pm}$.

There is an interesting connection between the result above and old speculations about the BMS group\cite{BMS} and its supersymmetric generalization\cite{awadagibbons}.  In effect we have shown that if we take  the large $N$ limit, in a Lorentz invariant manner, then the Hilbert space has operators corresponding to the angularly localized translations of the BMS group, and their SUSic generalization.  Even more interesting is the fact that translations appear as a consequence of SUSY.   We could imagine constructing a Lorentz invariant limiting theory even when there are no zero modes of the internal Dirac operator, and SUSY is broken, but the result would not be translation invariant.  Perhaps such limits exist and correspond to expanding ``bubble of nothing" space-times.

I see no reason for the S matrix to commute with all of these BMS generators.  They are the kinematic variables on which the S matrix acts.
More precisely, the conformal boundary of asymptotically flat space has two components, the past and the future, connected at space-like infinity.  The S-matrix maps the copy of the kinematic variables on the past boundary to that on the future boundary.  In a moment we will discuss the nature of the Hamiltonian for a geodesic observer, and argue that we can construct particle states, in a way that leads to asymptotic conservation of energy.  This argument does not lead to conservation of the angle dependent BMS generators.  The S-matrix maps the BMS generators on the past boundary into those on the future boundary.

Before proceeding to a discussion of particles, I want to note one other important feature of the HST description of spacetime.  It follows from the strong holographic principle that the {\it pixel superalgebra}
   $$[\psi_j (P), \psi_k (Q)]_+ = \delta_{jk} Z(P,Q) ,$$ has a finite dimensional unitary representation.  The classification of all such algebras and representations, is discrete, which means that {\it the parametrization of all possible compactifications in HST has no continuous moduli. }  This will come as a surprise to string theorists, who view the existence of continuous moduli as one of the aspects of string theory that has been established exactly.   In fact, in perturbative string theory, all we have established is that the dimensionful moduli appear to vary continuously in units of the string scale, but the string length scale is $\frac{1}{g_S^p}$ times the Planck scale so a Planck scale discreteness of the moduli is unobservable in the weak coupling limit.  If there is enough SUSY, one can prove ''non-perturbatively", in low energy effective field theory, that there is no potential generated on the moduli space.  But low energy effective field theory is an approximate formalism in which we have assumed the existence of a continuous moduli space, so using this as evidence for continuous moduli would be a circular argument. This example, by the way, suggests that conventional string theory attempts to fix the moduli in terms of non-perturbative potentials, cannot be viewed as an approximate implementation of the Planck scale discreteness that is alleged by HST.  The discreteness of moduli persists even in cases where no potentials are allowed.
   
The AdS/CFT correspondence provides another window on the question of moduli.  It's significant that when AdS space is realized as the near horizon limit of an extremal black brane in an asymptotically flat string compactification, most of the moduli are frozen by equations that depend only on the integer charges of the branes.  In the $AdS_3$ case, even the string coupling is quantized.  The only remaining moduli in AdS are parametrized by the conformal manifold of the CFT that defines the theory, which is typically a low dimensional space\cite{leighstrasslerseibergetal}.  
In $AdS_5 \times S^5$, the modulus is related to the ratio of the string tension to the Planck scale and is thus connected to the conventional picture that weakly coupled string theories are limits of theories of wrapped branes with quantized tensions, in which some dimensions of the compact manifold shrink to sizes smaller than the Planck scale, while the Planck length itself is scaled to zero, with the string tension fixed.  

In HST one must realize these limits in a somewhat different way, since the volume of the holographic screen in Planck units is always larger than ${\rm ln}\ 2$.  We are in the process of working this out for the two torus, and it appears that the resolution has to do with the fact that fuzzy spaces have only some of the topology of their continuous limits, and this allows for fractional values of brane charges (wrapping numbers) from the continuum point of view.  At any rate, this seems to be the key point in understanding the origin of continuous moduli in HST.

It is worth making one final remark about $AdS_d$ models in HST.  From the space-time point of view, the key feature of maximally symmetric spaces with negative c.c. is that the area of a causal diamond goes to infinity at finite proper time, of order the space-time radius of curvature.  The conformal boundary of diamonds of larger temporal extent is an interval times $S^{d-2}$, and this suggests a quantum theory whose symmetry group is the isometry group of $AdS$.  Given a finite system like the pixel algebra of a causal diamond, we are familiar with the way to extract limits invariant under the conformal group.  This is the technology of the renormalization group and CFT.  We can certainly find non-supersymmetric limits of this type.   However, if we insist that the c.c. be small in string units, which means that there is a parametrically large gap between a set of operators dual to particles in an almost flat bulk, and the generic, exponentially growing spectrum of operators in a typical CFT, then we expect the Mellin transformed correlators of the low dimension operators in the CFT to converge to a Poincare invariant S-matrix.  In HST this only seems possible with exact SUSY (in quantum effective field theory in four dimensions, it is only natural if we have exact SUSY and a discrete R symmetry).  Indeed, all known examples of CFT with a gap, actually have superconformal symmetry. 
Attempts to construct non-supersymmetric field theories with large radius AdS duals by orbifolding supersymmetric examples or constructing holographic renormalization group flows, all fail. In a model universe replete with many large families of super-conformal field theories with large radius duals, the difficulty of finding SUSY violating examples is remarkable and hints at the general principle we have adduced from HST.

\section{Particles, horizon states, and super-Poincare invariant limits}

We will concentrate on examples with $4$ non-compact space-time dimensions for two unrelated reasons.  The most important of these is the obvious relevance to the real world.  In addition, the analysis of the four dimensional case relies on well known results about the scaling of large $N$ matrix models.  In $d$ space time dimensions, the cutoff spinor bundle on the $d-2$ sphere resembles a rank $d-2$ anti-symmetric tensor of $U(N)$ or $O(N)$.  The classification of dominant large $N$ interactions for such variables is more complicated.  

On the two sphere, the fuzzy chiral spinor bundle is an $N \times N+1$ complex matrix $\psi_i^A (P)$ , and its conjugate is the corresponding $N + 1
\times N$ matrix $(\psi^{\dagger})_B^j (Q)$.  $P,Q$ label the sections of the cutoff spinor bundle on the compact internal manifold $K$.  The algebra
$$[\psi_i^A (P) , (\psi^{\dagger})_B^j (Q) ]_+ = \delta_i^j \delta_B^A Z(P,Q) ,$$ with commutation relations for $Z$ that form a closed, finite dimensional, unitary, fermionically generated representation of a super-algebra. This superalgebra representation characterizes the model and is the HST analog of a string theory compactification.  The dimension, for fixed $i,j,A,B$ of the representation is $e^L$ .  The radius of the 2-sphere is given by
$$ \pi (RM_P)^2 = N(N + 1) L  .$$  When the internal cutoff is large, the internal manifold has volume large compared to the Planck volume and $L$ is proportional to the volume in Planck units, thus reproducing the Kaluza-Klein scaling relating the $4$ dimensional Planck scale to the higher dimensional scale.  This works because the spinor bundle of a product manifold is the tensor product of the individual spinor bundles.

Consider a state satisfying  $$\psi_i^A (P) | {\rm particle} \rangle = 0$$ for $KN + Q$ matrix elements.  Think of this as a constraint on initial states. Here $N \gg 1$ is the proper time difference between the past and future tips of a large causal diamond in Minkowski space.  We will be interested in the regime $ 1 \ll K,Q \ll N$.  The $N \times N$ matrices, $M$, are defined by $$M_i^j \equiv \psi_i^A  (\psi^{\dagger})_A^j $$.  Acting on these states, they become block diagonal, with $K_i \times K_i$ blocks satisfying
$$ \sum K_i = K ; \ \ \ \ \ \sum_{i \neq j} K_i K_j = Q .$$ There is one large $(N - K) \times (N - K)$ block, which we call {\it the horizon DOF}\footnote{Actually, these should be called the {\it active horizon DOF}.  The frozen DOF are also associated to the horizon - they're certainly not particles.}. 

The constraint defines a subspace of the Hilbert space, as well as a tensor factorization of the subspace into DOF associated with each block.  On this subspace we will choose a time dependent Hamiltonian
$$H = H(N) = P_0 + \frac{1}{N^2} {\rm Tr} \sum_{k=2}^R g_k (N) M^k , $$ with $R$ a fixed integer, independent of the variable $N$.  $P_0$ is a Hamiltonian, to be specified below, which commutes with the constraint on states.  It is bilinear in the fermionic variables.  We will use a time dependent Hamiltonian of this form for all values of $N$, not only asymptotically large ones.  The coefficients $g_k$ can be functions of $N$, which approach constants at large $N$.  Recall also that $N \geq 1$, so the Hamiltonian is not singular for any value of $N$.  We will refer to the small blocks as the particle DOF, while the rest of the variables are called the horizon DOF.  
The constraint on asymptotic states is such that particle and horizon DOF are decoupled in the large $N$ limit.

The higher order terms in the Hamiltonian violate the constraint, but because $P$ is fixed, they cannot change $K$.  The terms in the Hamiltonian involving the DOF that vanish on the initial state, must act, and they can act at most $P$ times with one application of the Hamiltonian.  Note that the scaling of the interaction term has an additional factor of $1/N$ relative to the standard large $N$ matrix scaling, so the interaction is a small perturbation of $P_0$ for large $N$.  Recall also that for smaller values of proper time most of the DOF in the large causal diamond, simply do not appear in the Hamiltonian describing the smaller causal diamond.  The terms which {\it do} act on these outside variables at smaller proper time, are constrained only by consistency with the Hamiltonians of other observers, on overlaps.  We will present a preliminary discussion of those constraints below.
We conclude that, for this entire class of Hamiltonians, the integer $K$ appearing in the constraint on incoming states, is an asymptotically conserved quantum number.  $Q$ is not however conserved, and will turn into another integer, $Q^{\prime}$.  

We can of course take the initial state to be a tensor product of a state of particles and a state of horizon DOF, since that is a basis of the Hilbert space for any fixed value of the constraint parameter $K$.  We now want to argue that the final state will also be a tensor product of the new particle state, characterized by $Q^{\prime} $ and the values of the individual $K_i^{\prime}$, and a state of the horizon DOF.  To do so, take $N= N_0$ large enough so that $Q^{\prime}$ no longer changes.  Now consider time evolution for times larger than $N_0$ but with the Hamiltonian taken to be constant and equal to $H(N_0)$. This time evolution operator will produce {\it more} entanglement between particles and the horizon than the actual time dependent Hamiltonian does, because the time dependent Hamiltonian turns off the coupling between the two subsystems as time goes to infinity.

Sekino and Susskind\cite{sekinosusskind} have conjectured that generic large N matrix Hamiltonians are {\it fast scramblers}.  While I think this conjecture is very plausible, it is more than we need to assume.  It is sufficient that there be a range of choices for the $g_k$ such that the Hamiltonian $H(N_0)$ is a fast scrambler.  If that is the case then, in a time of order $N_0$ the state of the system will be thermalized.  The time averaged density matrix will be maximally uncertain and will have entropy of order $L N_0^2 $.  The probability of finding a tensor product particle state, satisfying the constraint  that $N_0 K + Q^{\prime}$ matrix elements of $\psi $ vanish 
is of order $$P(K) = e^{- N_0 L K} .$$   This is a thermal state, if we think of the asymptotically conserved quantum number $K$ as a dimensionless measure of energy.

We can think of the real time evolution of the system by simply following the above procedure, but taking $N_0$ larger and larger.  Thus, the entanglement between particles and horizon, which is smaller than that implied in the thermal system described above, goes to zero with time.  We've already argued that the constraint of $NK + Q^{\prime}$ vanishing matrix elements is rigorously obeyed by the actual time evolution operator as the proper time $N$ goes to infinity.  Since $K$ is asymptotically conserved, the state cannot completely thermalize at zero temperature, as it would if we turned off the time dependence of the Hamiltonian, but the entanglement between particles and horizon does go to zero.   Essentially this is because the particle factor in the tensor factorization ${\cal H} = {\cal H}_{particle} \otimes {\cal H}_{horizon}$ only becomes a factor in the limit $N\rightarrow\infty$.  Particles and horizon DOF {\it do} mix at finite times, and the identification of which DOF to call particles and which to call horizon takes place only when $N$ is large enough for the separation to be clear.

The upshot of this analysis is that there is a non-trivial S-matrix for interaction between particles in this system, but that the interaction is mediated by horizon degrees of freedom, which are present at finite times, but decouple from the particles as the proper time goes to infinity.  More precisely, when $N$ is not that large, it becomes hard to distinguish between particles and horizon DOF, and they interact strongly.  The state that emerges from the future boundary of the causal diamond as $N$ goes to infinity, has a clean separation between particles and horizon, but it is not the same separation that was present in the initial state.  The number of particles, as well as their individual energies, changes. As we will see, the Hamiltonian $P_0$ describes only free particle propagation.  {\it The interactions are approximately local, not because we are integrating out high energy localized DOF, but because the interaction is constrained to take place in small causal diamonds, by the time dependence of the Hamiltonian. In fact, the horizon DOF have exactly zero particle energy, $K$, and the asymptotic Hamiltonian vanishes on all those states.}  

We've been using the word particle to describe the DOF in the small blocks.  The Hamiltonian we have described has an $S_k$ gauge symmetry, which permutes the particle blocks.  This is a subgroup of a large $U(N^2)$ symmetry of the interaction terms in the Hamiltonian.  The anti-commutation relations of the fundamental fermions $\psi_i^A (P)$, and the fact that they carry half integral spin under $SO(3)$ insures that this is the correct statistical symmetry for Bose and Fermi particles, with the right connection between spin and statistics. Another way of thinking about this statistical symmetry is that it is a holographic version of the usual field theoretic explanation of statistics.  Our variables are, in the large $N$ limit, a spinor field on the holographic screen, and particles are just localized excitations of this field, defined by the value of the local field, but otherwise indistinguishable.

Furthermore, if, in each $K_i \times K_i$ block,  we use the construction of SUSY generators described in the previous section, we obtain the algebra of operators of a single super-multiplet of particles.  The full representation space of the pixel algebra breaks up into a direct sum of such super-multiplets.   The energy operator $P_0$ in the SUSY algebra is proportional to $K_i$.   {\it Thus, the asymptotically conserved quantum number $K$ is just proportional to the sum of single particle energies, as defined by the SUSY algebras in individual blocks.}

So far, nothing we have said guarantees that the spatial momentum in the SUSY algebra is conserved.  This is a consequence of the fundamental consistency condition of HST.  Indeed, we can take our lattice of trajectories in Minkowski space to be geodesics separated by a rigid translation in space.  In the limit $N\rightarrow\infty$, the overlap between the causal diamonds along any pair of trajectories does not become the full Hilbert space, though the fraction of the entropy not contained in the overlap is $o(1/N)$ of the total entropy.  

However, I claim that for the particle subfactor, the overlap is complete.  The point is that in order to be a particle, a state must leave a ``track" in all causal diamonds larger than some value $N_{max}^i$.   Its longitudinal momentum and angular wave function should be, to a good approximation preserved for all $N$ larger than $N_{max}^i $.  For a finite number of particles there will be some maximal value of $N_{max}^i$.  

Even if there are IR divergences, so that outgoing states contain classical brehmstrahlung beams, we can recall that our variables are really super-BMS generators, which measure local flows (jets) of quantum numbers at infinity, and not necessarily isolated single particles (I'll comment on this more, in a moment).  We conclude that for any given S-matrix process, we can take $N$ large enough that, even though some jets contained in the causal diamond of trajectory $1$, are outside the causal diamond of trajectory $2$ at time $N$, {\it they have passed through the holographic screen of diamond 2, as freely moving jets, and thus affect the state in the particle sub-factor of diamond 2.}   The particle subfactor of the Hilbert space of trajectory 1 coincides with that of trajectory 2 as $N \rightarrow\infty$. 

Thus, the density matrix constraint becomes the statement that the pure states of any two trajectories must be the same in the asymptotic limit, which means that the S-matrix for spatially translated trajectories, is invariant under spatial translation.  

A similar argument, applied to geodesics which have constant velocity w.r.t. each other, implies that the S-matrix is invariant under boosts.   While, from a purely group theoretical point of view, there are two possible symmetry groups, Poincare or Galilean, the fact that HST incorporates Einstein causality in an explicit manner, combined with the explicit construction of the asymptotic  Hamiltonian in terms of SUSY generators implies that the asymptotic symmetry group must be the super-Poincare group.  This is of course consistent with the fact that the dispersion relations for the particle states we have constructed are Lorentz invariant.  

We emphasize that we are not claiming that asymptotic super-Poincare invariance follows automatically, as time translation invariance did,  for any choice of the $g_k$ above.  Rather, the density matrix constraints of HST imply that the $g_k (N)$ must be chosen, for large $N$, in such a way that the S-matrix is super-Poincare invariant.  We have not proven that this is possible, but since the class of Hamiltonians we study contains many free parameters for large $N$, we believe that it is plausible.  

I'd also like to note a constraint on the pixel algebra, which follows from these considerations.  We've seen that, given a covariantly constant spinor on the internal manifold, we get massless superparticles\footnote{The generalization to massive BPS particles for toroidal compactifications is less straightforward than might have been imagined, but appears to work.}.  The full pixel algebra must be such that, under the SUSY subalgebra, there is exactly one superparticle of maximal spin $2$, and none of higher spin.  The problem of classifying all such super-algebras is completely open, and I hope that someone will take it up.  It is, of course, a purely kinematic exercise, analogous to classifying consistent free super-string theories.  If string theory is a guide, many of the kinematic solutions will not be compatible with a non-trivial unitary, super-Poincare invariant S-matrix.

Having spent some effort in trying to convince you that a class of HST models contains particles, I now want to confess that this is not precisely correct.  Our analysis so far identifies excitations on a holographic screen at infinite proper time $N$, and shows that they carry spin and momentum.  However, since we are only talking about the holographic screen, a better name for these excitations is ``jet detector elements", or pixels.  The pixels have an angular width of order $\frac{1}{K_i}$, and a total longitudinal momentum of order $K_i$\footnote{For fixed $N$, $K_i$ is roughly the longitudinal momentum in Planck units, if the internal dimensions are Planck scale.  In the large $N$ limit, only ratios of $K_i$ are important.} There is no reference at all to the time delay between different events, which deposit energy in the same pixel.  This is as it should be, because time delays are encoded in phases in the S-matrix, and our discussion has been purely kinematic.  In HST, a particle state will be defined not just by the asymptotics, but by the constraint that we be able to follow the particle for some range of $N$, thus establishing a localized ``track" in the pixel detector variables.   Even then, our variables do not really distinguish states which differ only by what a field theorist would call the addition of massless, almost collinear, particles.   This is a good thing, and I expect it will lead to a definition of the gravitational S-matrix that is always free of IR divergences.

Thus, our variables really detect all possible multi-particle states, or jets, with total longitudinal momentum proportional to at most $K_i$ and angular resolution $\frac{1}{K_i}$.   The smallest longitudinal momentum for particles describe by this block of variables, at fixed $N$, is $\frac{K_i}{N}$, which is the same as the maximal transverse momentum.

I suspect that this means that the HST formalism will be free of IR divergences in $4$ space time dimensions.  It will not compute the S-matrix for single particles, but the amplitude for a certain amount of localized energy (and spin) to come in through particular pixels on the past conformal boundary of Minkowski space, and exit via some other set of pixels on the future conformal boundary.  Even in QCD perturbation theory, observables like this are IR safe, and the milder infrared divergences of gravity should also disappear.   I would imagine that there is some connection to the IR finite matrix elements defined by Kulish, following the classic work of Fadeev and Kulish\cite{fadkul} on QED, but I do not know the connection.

These observations are also relevant to our discussion of BMS generators.  The most careful analysis of the BMS group that I know, was given by Ashtekar in a little book called {\it Asymptotic Quantization}\cite{ashtekar}.  The relativist's insistence on the BMS group, rather than the Poincare group, stems from the desire to classify space-times with classical ingoing and outgoing gravitational radiation as ``asymptotically flat".  One can impose stronger boundary conditions, which define what Ashtekar calls {\it the vacuum sector},  and have only the Poincare generators as infinitesimal asymptotic isometries.   Ashtekar, working in four dimensions, argues that one cannot restrict attention to {\it the vacuum sector} precisely because of gravitational brehmstrahlung. In higher dimensions, an S-matrix in which finite numbers of ingoing particles produce finite numbers of outgoing particles is well defined, and there is no need to consider classical gravitational radiation, except as an approximation in which a large finite number of bosons is viewed as a classical field.
I believe that the HST formalism provides an important supplement to these ideas.  The natural variables are localized distributions of energy and spin on the holoscreen, and amplitudes for these should be well defined in any dimension.  The angle dependent translations of the BMS group, are the variables that define these distributions, rather than generators of asymptotic symmetries .  They are the natural language for thinking about Ward identities, which relate amplitudes with different numbers of outgoing jets of energy, when some of the particles in the jets are soft gravitons\cite{strominger}. The asymptotic Hilbert space should be defined as a representation of the super-BMS algebra, rather than a Fock space of particles.

\section{The Theory of Stable dS Space}

Our discussion of finite causal diamonds in Minkowski space, leads directly to a conjecture for a theory of stable dS space.
Indeed, the Hamiltonian for that system, along a timelike geodesic, is simply the Hamiltonian for the corresponding Minkowski system, up to proper time $N_{max}$ at which the area of the causal diamond is equal to the area of a maximal causal diamond in dS space. From that time forward, we propagate the system with the time independent Hamiltonian $H (N_{max} )$,  operating on a finite dimensional Hilbert space with entropy of order $N_{max}^2$. 

This system has states, which, for times $< N_{max}$ behave in a manner identical to particles in a finite causal diamond in Minkowski space.  For times $> N_{max}$ the approximate decoupling between particles and horizon states begins to break down, and the system thermalizes.  Averaged over a relaxation time which is of order $N_{max} {\rm ln}\ N_{max}$
if the terms $\sum g_k {\rm Tr}\ M^k$ have the fast scrambling property, the density matrix becomes the maximally uncertain density matrix on the Hilbert space.   The probability of finding a state with a jet of angular localization $1/K$ is
$e^{- K N_{max} \sqrt{L}} $ \footnote{Recall that $L$ is the entropy of the single pixel Hilbert space, which counts excitations of the compact dimensions.}, which is a thermal distribution since $K$ is proportional to the energy of the jet.  More precisely we have
$$L N_{max}^2 = \pi (RM_P)^2 ,$$ so that $$N_{max} =\sqrt{\frac{\pi}{L}} RM_P . $$     This will be the standard semi-classical thermal distribution in dS space if $$K = 2\sqrt{\frac{\pi}{L}} \frac{E}{M_P} .$$

The picture that emerges is that a stable dS space is a thermal system, with the temperature (for localized fluctuations) and entropy indicated by the semi-classical considerations of Gibbons and Hawking\cite{gibbhawk}.  It has localized excitations, which behave like elementary particles in a single horizon volume, or like black holes.  All of these excitations ``decay" into the dS vacuum, which is simply the ensemble of typical states in the full Hilbert space.  The fastest decays are simply the geodesic motion of local excitations which are not bound to the origin of the local static coordinate system.  These excitations have lifetimes of order $R$.  Black holes sitting at the origin have lifetimes of order $M^3$, when their mass is much less than the maximal (Nariai) black hole mass in dS space.   Excitations of charge $Q$ sitting at the origin have lifetimes of order $R e^{2\pi Q m_e R}$, where $m_e$ is the mass of the lightest charged particle in the model, and $Q$ is the charge in units of the lightest particle charge.  The mechanism of decay is the thermal nucleation of $-Q$ units of charge at the origin, its annihilation with the existing charge, and propagation of the annihilation products out to the horizon.
We see that, while the life-times of localized excitations can vary widely, there are no stable localized excitations in dS space.

This is consistent with the fundamental observation that any localized object sitting at the origin has less entropy than the empty dS ``vacuum".  Since the object's trajectory is time-like, it must have mass, $M$, which means that its faraway gravitational field must be the dS-Schwarzschild metric (or a metric with charge and angular momentum).  The black hole horizon may of course be too small to be considered part of the far field, but the shift in the dS horizon certainly is not, and leads to a decrease in entropy $$\Delta S = 2\pi M R ,$$ for $M \ll R$ in Planck units.  In eternal dS space, the only way to create localized excitations is through thermal fluctuations.  Dyson-Kleban and Susskind\cite{DKS} have used this to argue that eternal dS space is not an acceptable model of our universe.  The basic point is that we have evidence for vast amounts of cosmic history and vast reaches of space populated by galaxies full of stars.  A fluctuation that created only a tiny subset of those localized excitations, is all that is needed for the existence of observers, and so the real history of the universe is doubly exponentially less probable in this model, than an alternative one in which we could still observe, but our observations would be different from what we see.  This argument goes back to Boltzmann, and his assistant Schutz.
We will see that the HST model of cosmology does not suffer from this problem.

The conventional global picture of dS space, is implemented in HST by constructing a cubic lattice of trajectories with identical Hamiltonians.  The overlaps between the causal diamonds of these trajectories are imposed by referring to causal diamonds associated with fixed time slices in the flat slicing of dS space.   The lattice defines the geometry (and not just the topology) on the $t=0$ slice, and has Planck scale spacing there. Any two points, a fixed number of Planck units apart on that slice, become causally disconnected after a finite proper time and the overlap conditions between them become trivial.  

I will end this section on dS space by reviewing the connection between the c.c. and the breaking of SUSY, which was first hypothesized in \cite{susyholo}.  We've seen that in HST, interactions between particles are mediated by horizon DOF.  For a given choice of incoming energies, it becomes impossible to distinguish between particles and horizon excitations below a certain size of causal diamond.  It's the time dependence of the Hamiltonian that forces the restriction of the interaction to small causal diamonds, which is responsible for the approximate locality of particle interactions.  Once one comes to the future boundary of a large causal diamond, the state can once again be interpreted as a product state of particle and horizon DOF, but the number of particles, and their individual energies and momenta, have changed.  We've hypothesized that it's possible to choose the interaction coefficients, $g_k$, in such a way that the scattering amplitudes are Lorentz invariant, and argued that this {\it must} be done in order to satisfy the consistency conditions of HST. 

Ancient results\cite{mandelwein} imply that when the kinematic invariants $p_i \cdot p_j$ are small compared to the Planck scale\footnote{The Planck scale and the discrete parameters labeling the pixel algebra and its representations are the only parameters in HST.  If the internal dimensions are Planck scale, {\it i.e.} the discrete parameters are order one, the Planck scale is the only scale that appears in amplitudes.}, then the S-matrix must be equal to one computable in QUEFT.  We are interested in possibilities in which the relevant QUEFT is $d=4$ minimal SUGRA, coupled to a number of chiral and gauge multiplets.
For such models, vanishing c.c. is automatic, only if there is both SUSY and a discrete complex R symmetry.  Furthermore, there is (generically) no continuous moduli space if the number of R charge 2 chiral fields is greater than or equal to the number of R charge zero fields.   Since our underlying HST model of stable dS space is unique, we would expect that it leads to an isolated QUEFT, with SUSY and a discrete complex R symmetry, in the limit of vanishing c.c. .

The corrections to the effective Lagrangian of this QUEFT, coming from a small but non-vanishing c.c. should be computable, to leading order, in terms of particles of the theory with vanishing c.c., interacting with the horizon variables on the finite area horizon.  The Lagrangian must lead to SUSY violation, because there are no unitary representations of the $SO(1,4)$ covariant SUSY algebra. In SUGRA, as in any gauge theory, violation of SUSY can always be thought of as spontaneous.  Moreover, since we can always dial the scale of SUSY violation to zero, by choice of the c.c., the Goldstino must appear as a component of an unconstrained chiral or gauge supermultiplet.  

We might expect that the Lagrangian corresponding to vanishing c.c. has generic couplings consistent with its symmetries.  Turning on the c.c. will generate corrections to these couplings, which should vanish like a power of $\Lambda$ in Planck units, and are therefore negligible unless the corresponding terms vanish identically, which means that they break a symmetry.  The complex R symmetry is the obvious candidate, since the gravitino mass term has R charge 2 and breaks any R symmetry to $(- 1)^F$.  

If we think in terms of Feynman diagrams, as experience with string theory in asymptotically flat and AdS space leads us to do, then the diagrams which break R symmetry, once we put in the dS horizon, are diagrams in which a gravitino propagates out to the horizon, and another is emitted from the horizon and comes back to the vicinity where the gravitino began (thus generating a correction to the gravitino propagator between two points near the origin), which breaks R symmetry.  If the gravitino has a mass, such diagrams are suppressed by a factor 
$$ e^{ - 2 m_{3/2} R},$$ because of the space-like separation between the horizon and the points between which the real gravitino propagates.  This diagram violates R symmetry by two units (since the gravitino momentum is reversed but its spin unchanged).  Note that even if the gravitino mass is small in the limit $R \rightarrow\infty$, we cannot neglect it in this formula because it is multiplied by $R$.  

Quantum mechanical perturbation theory tells us that the interaction of the gravitino with the horizon, which gives rise to a non-zero value for this diagram, is of the form

$$ V^{\dagger} \frac{1}{\Delta E} V ,$$ where $V$ is an operator corresponding to the absorption of the gravitino by the horizon and $V^{\dagger}$ the corresponding emission operator. The typical value of $\Delta E$ is a redshifted value for the gravitino energy, which is a power of $R$.  There is however an exponentially large number of states on the horizon, namely $e^{\pi (RM_P)^2}$.  The gravitino, since it is a massive particle, cannot propagate on the horizon for a proper time longer than $\frac{1}{m_{3/2}}$.  Given the standard formula for the propagator, this means that it can cover an area of order 
$\frac{1}{M_P m_{3/2}}$ in this amount of proper time (our underlying model has a Planck scale UV cutoff on proper time).  Thus, we expect the second order perturbation theory formula gives a contribution of order $$ e^{\frac{c M_P}{ m_{3/2}}} .$$  This formula is accurate to exponential order in powers of $R$, and has power law corrections.  

As a consequence, we expect that R symmetry violating interactions in the effective theory of particles in dS space are of order
$$g_{\Delta R} \sim e^{ - 2 m_{3/2} R} e^{\frac{c M_P}{m_{3/2}}} R^p,$$ and this should be true of the gravitino mass itself.
Now notice that if $m_{3/2}$ vanishes more rapidly than $R^{ - \frac{1}{2}}$ as $ R\rightarrow\infty$, then $g_{\Delta R}$ blows up, contradicting the hypothesis that the limit of Minkowski space is smooth.  On the other hand, if we assume $m_{3/2}$ vanishes less rapidly than $R^{ - \frac{1}{2}}$ then all R violating interactions, including the gravitino mass vanish exponentially, which is again a contradiction.  The conclusion is that, up to corrections that are sub-leading in $R$, $m_{3/2} \sim R^{ - \frac{1}{2}}$.  This formula is written in Planck units, and is equivalent to $m_{3/2} \sim \Lambda^{\frac{1}{4}} = 10^{-3}$ eV.  More refined considerations suggest that this should be multiplied by $(M_P R_I)^{1/2}$, where $R_I$ is the scale of the internal dimensions.  If that scale is identified with the scale of coupling unification, then we get an extra factor of $10$ in the gravitino mass.  Even so, this is an extraordinarily low scale for the largest breaking of SUSY in the particle spectrum, of order $F \sim 30 ({\rm TeV})^2 $.  This has important consequences for phenomenology, and the most general model compatible with these ideas is already in tension with the data\cite{pyramid}.  We have no space to go into detail here.

In a conventional QUEFT treatment of the relation between the gravitino mass and the c.c., one argues that the ``natural" value is $m_{3/2} \sim \sqrt{\Lambda}/M_P = \frac{1}{R}$. In HST a particle with momentum this low, would have no angular localization at all, and could not be considered a particle.  Furthermore, when plugged into our formula for R symmetry violating interactions, this gravitino mass would give a divergent result in the supposedly smooth limit of infinite dS radius.  The way out of course is that QUEFT does not really make a prediction for the c.c. expected for a given amount of SUSY breaking.  One can always tune the c.c. by manipulating the R symmetry violating constant in the superpotential, which is unobservable in the absence of gravity.  From the QUEFT point of view, this is unnatural fine tuning.  However, string theory has only taught us that the rules of naturalness in QUEFT apply to parameters in the QUEFT which approximates the boundary correlators with a fixed asymptotic geometry.  By contrast, everything we know about the c.c. from AdS/CFT suggests that is a fixed parameter, which determines the point in proper time at which the area of a causal diamond goes to infinity.  In CFT terms, it is determined by the high temperature value of the free energy, and is a parameter in the UV definition of the CFT, not some calculable low energy field theory parameter.

Similarly, the conjecture of Fischler and Banks\cite{tbwf} is that positive c.c. determines the point of fixed area at which proper time goes to infinity.  Quantum mechanically, it is the entropy of the Hilbert space describing stable dS space. It does not have the same status as other parameters in low energy field theory.  
More generally, the fact that parameters violating the discrete R symmetry of the zero c.c. limit come from interaction with degrees of freedom on the dS horizon, and involve very special graphs in QUEFT, means that none of these parameters have to satisfy the usual rules of naturalness in QUEFT.  This can have important phenomenological implications and may for example be the correct solution of the strong CP problem.
The rules for calculating these parameters are not yet clear, but they will involve apparent fine tuning in QUEFT, which simply implements relations derived from the underlying theory.

I want to make one further point about stable dS space before turning to the discussion of holographic cosmology.  We described the local physics of the model by comparing it to the limiting model with vanishing c.c., which was supersymmetric.
In higher dimensions, there are simply no supergravity theories, which arise from compactification of 11D SUGRa, which have spontaneously broken SUSY and dS solutions. This suggests that the quantum theory of stable dS space only exists in four space-time dimensions.

\subsection{Accelerated Observers}

HST suggests that there should be a time dependent Hamiltonian for {\it any} time-like trajectory in any Lorentzian space-time which describes the thermodynamics of a real quantum gravity system.  In particular, in any causal diamond in Minkowski space there exists a sequence of uniformly accelerated trajectories that share the same past and future tips.  These are lines on time-like hyperbolae of different curvature, but with different foci.  In the limit as the proper time in the diamond goes to infinity, they become the trajectories studied by Unruh, which are flow lines of the vector field representing a Lorentz boost $J_{0i}$.  Similarly, in dS space, there are lines of fixed radius and fixed direction $n_i$ on the two sphere.  Our proposal for the Hamiltonian for such trajectories in a diamond of proper time $N$ is
$$ H_i (a;r) = \sigma_i Z(a;r) \sum_k P_0^{(k)} + \frac{1}{N^2} {\rm tr}\ \sigma_i f(\psi \psi^{\dagger}) . $$ Here the semicolon notation indicates the fact that this describes either an accelerated Minkowski trajectory at fixed acceleration, or a 
dS trajectory at fixed $r$.  $\sigma_i$ is an operator built from the $\psi$ variables which describes a single q-bit with angular momentum generators $\frac{\sigma_i}{2} $.  The renormalization factor $Z(a;r)$ implements the redshift of energies familiar from the Rindler metric or the static coordinate frame on dS space.  It varies between $1$ for the geodesic observer and $\frac{1}{N}$ for the maximally accelerated observer. The locus of all maximally accelerated observers defines the stretched horizon for these systems. At this value of acceleration the part of the Hamiltonian that implements free particle motion is no larger than the interaction terms, and for these observers the system is completely thermalized, even when $N$ is large.  Note that the maximally accelerated Rindler observers play a crucial role in Jacobson's derivation of Einstein's equations from thermodynamics.  Our definition of these trajectories obeys the assumptions that went into Jacobson's argument.

\section{Holographic Cosmology - Three Simple Models}

Holographic Cosmology starts out with a bang.  A Big Bang, which turns out to be more of a whimper.   Indeed the mere postulate that the universe starts out at some finite proper time, along some trajectory, means that the causal diamonds going back to that initial time, get smaller and smaller as the final proper time approaches the initial point.  So, from the point of view of that trajectory, the Hilbert space describing the interior of the diamond shrinks, and its quantum mechanics becomes that of a single q-bit but does not become singular.  A natural hypothesis about the very early universe is that it is highly excited.  The state inside a causal diamond should jump randomly.  We can implement this by postulating a random choice of Hamiltonian at each step, chosen from some ensemble, and involving all of the pixel variables available at that point.  

So far, we have only explored cosmologies in which the pixel algebra remains constant for all time, corresponding to a fixed structure of compact dimensions.  We also restrict attention to 4 non-compact dimensions, though it's fairly easy to generalize our first model to any number of dimensions. 
Finally, we will assume that the Big Bang hypersurface has the topology of infinite flat 3 dimensional space.  A positively curved initial surface would generically lead to a Big Crunch in which the maximal sized causal diamond was small.  

The time step in our model is taken to be the Planck time.  This is done for convenience.  In general, if one adds precisely one pixel algebra worth of degrees of freedom at each step, then the proper time difference goes to zero because the area increase goes like $(\Delta t)^2$ in four dimensions.  However, there is no way to add a single pixel in a rotation covariant way, because an angular momentum multiplet has $2L + 1$ components.  The overlap rules for such tiny time steps would have to be correlated with the choice of direction in which the new pixel was added.  To simplify things we choose to add a full angular momentum multiplet at each step, simply taking $N \rightarrow N + 1$ for our cutoff spinor bundle.   This procedure removes a problem with the original version of holographic cosmology\cite{bfmholocosm}.  
It is compatible with simple overlap rules.  It is probably possible to make a more fine grained time resolution, with more complicated rules, which would reduce to our simple rules every time a full Planck interval had passed.  We have not been able to identify any observational consequences of such a refinement, and so have not pursued it. The model is designed to saturate the Covariant Entropy Bound at all times.
In their seminal paper, Fischler and Susskind\cite{fs} showed that generic cosmologies violate the bound at early times and that the flat FRW cosmology with equation of state $p =\rho$ can saturate it.  We will see that for large causal diamonds the thermodynamics of our model is exactly described by the $p = \rho$ FRW.  

As usual, the model has of order $L N^2$ active DOF, when the proper time along the trajectory is $N$ Planck units.  We will insist that, as $N\rightarrow\infty$, the Hamiltonian approaches that of a $1 + 1$ dimensional CFT with central charge of order $N^2$, living on an interval of length $N$, with a UV cutoff of order $1/N$.   The density matrix is maximally uncertain so the entropy is of order $N^2$, while the average energy is of order $N$.  Note that in order to use these CFT estimates the interval must be much larger than the UV cutoff.  This occurs if $L$ is large, which is an amusing connection between the size of compact dimensions and early universe cosmology.  Frankly, I do not know what to make of it.  

The CFT can be perturbed by a random power law irrelevant operator.  There is an interesting question in CFT raised by this: if the CFT is integrable, is the random perturbation in the UV enough to thermalize the system?  In our original papers on this cosmology we treated the variables as fermions and used the observation\cite{kapwein} that a random bilinear in fermions gives the massless Dirac equation in the large $N$ limit.   We assumed that the random irrelevant perturbation was enough to thermalize the system.   It may be that one must use a non-integrable CFT instead.  

Although we will not pursue this here, there is an interesting interpretation of the space-time of the CFT in terms of the picture of this cosmology as a ``Dense Black Hole Fluid".  In that picture one claims that to an ``external observer", the particle horizon of the $p = \rho$ universe looks like the horizon of a black hole.  The degrees of freedom are associated with the horizon, but the geometry of the horizon itself is described by the spinor variables.  The interval of the CFT may be interpreted as a space-like coordinate transverse to the horizon and stretching of order a horizon distance from it.  There might be some connection between this picture and attempts\cite{carlip} to use Cardy's formula to understand the entropy of generic black hole horizons.

Returning to the emergent space-time geometry of our model, it is fixed by giving overlap rules for an arbitrary trajectory in a 3 dimensional cubic lattice on the Big Bang hypersurface.  Given two points ${\bf x,y}$ on the lattice, we denote the minimal number of lattice steps between them by $d({\bf x,y})$.  Our rule for the overlap is that the overlap Hilbert space is $${\cal O} ({\bf x,y} ; N) = {\cal H} (N - d({\bf x, y}) ; {\bf x}) = {\cal H} (N - d({\bf x,y}) ; {\bf y}).$$ That is, the overlap is the Hilbert space encountered along each trajectory $d({\bf x, y})$ steps in the past.  If we give the same dynamics, for states both interior and exterior to the causal diamond, for all trajectories, then all overlap conditions are satisfied.  Note that we do not have to specify what the evolution operator is outside the causal diamond, only that both it and the initial state are the same for all trajectories.  This feature of the $p=\rho$ cosmology is undoubtedly due to its {\it lack} of features, by which we mean local excitations.   

In the space time geometry defined by the causal structure implicit in the overlap conditions, all points that are the same number of lattice steps away from a given point, have the same spatial distance from it, as given by the light travel time.  The locus of all such points is a cube tilted at $45$ degrees, so the geometrical shape of this cube is a ball with a round two sphere as its boundary.  This inspires us to choose our Hamiltonians to be exactly $SU(2)$ invariant, which we are able to do because we always work in Hilbert spaces that are representations of $SU(2)$.  

We see that, although we have chosen random Hamiltonians and generic initial states, a homogeneous isotropic universe describes the coarse grained geometry on scales larger than the Planck scale.  The volume of space in this Jacobsonian geometry is of order $N^3$ at proper time $N$. The energy and entropy densities are thus
    $$\rho \sim \frac{1}{N^2} ,$$ $$ \sigma = \sqrt{\rho} ,$$ which are the thermodynamic relations for the equation of state $p = \rho$.  The time dependence of these quantities follows the Friedmann equation for a spatially flat universe.  This is a special case of Jacobson's general result that the thermodynamics of a system satisfying the Bekenstein Hawking law will be give by Einstein's equation.  Once we've derived these relations, we can invent a classical scalar field $\phi$ with no potential, and view the equations as devolving from a Lagrangian for gravity coupled to this scalar.  It would be a mistake however to conclude that quantized excitations of the scalar or classical inhomogeneities in it are related to the actual quantum physics of the model.
 THEFT rules but QUEFT is irrelevant, in the early universe\footnote{We also note that the BKL\cite{bkl} analysis of Einstein's equations near a space-like singularity, reveals the classical theory to be in a kinetic dominated regime, with random variations in the fields from point to point.  This has the same equation of state, $p=\rho$ as our quantum model, but the infinity temperature entropy of such a system of spatially decoupled classical fields is too large to satisfy the Covariant Entropy Bound.  Furthermore, our general considerations have shown that the assignment of DOF to points interior to a causal diamond, rather than to its holographic screen, makes sense only in {\it low} entropy situations. and even there, it neglects most of the DOF, which are decoupled from the localized excitations.}.   
 
 We should also note that the Big Bang is a phantom in this model.  The scaling regime where the emergent geometry is a good coarse grained description is, like most models of an asymptotic future scaling regime, singular when extrapolated into the past, but does not describe the real past of the model.
 In the past the system inside the horizon simply has too few degrees of freedom to justify a thermodynamic approximation, but the true dynamics is non-singular.  Quantum gravity avoids the singularity not with a bounce, or a transition to some other pre Big Bang behavior, but with a whimper.  Backwards evolution simply runs out of steam because the Hamiltonian inside the causal diamond has no more states to act on. 
 
 There is a simple variation on this model cosmology in which we simply stop the growth of the Hilbert space at some finite value $N_{max}$ and let the system evolve with the Hamiltonian $H( N_{max})$.  Actually, we conformally transform the $1 + 1$ dimensional theory to one living on an interval $\sim N^3$ with cutoff $1/N^3$.  The expectation value of the energy is now $o(1/N)$, which coincides with the energy scales in a dS space of radius $N$. The causal structure is also that of dS space, proper time goes to infinity while area stays finite.  Trajectories that are more than $N_{max}$ lattice steps apart, never have overlapping causal diamonds.
 
 After the rescaling, the Hamiltonian has the properties we've attributed to the Hamiltonian of the trajectory at the stretched horizon in dS space. The approximate $1 + 1$ dimensional scale invariance is surprising in the dS context, but over time scales that are larger than R the system starts to explore much of its Hilbert space, where the neglect of the irrelevant perturbations may be unjustified.  On the other hand, we have already speculated on the possible relevance of $1 + 1$ dimensional conformal invariance to generic horizons, so perhaps there is something to be learned here.
 
 What is completely absent in the dS space that arises in the asymptotic future of this model is any hint of localized excitations and Poincare invariance.  This is a consequence of the fact that at early times we have taken the initial state to be quite different from the constrained scattering states of particles in our model of Minkowski space.  Furthermore, the time dependent Hamiltonian is designed to be such that, at each point in time, the state of the system is a random tensor product state in the ${\cal H}_{in} (n)\otimes {\cal H}_{out} (n) $ Hilbert space.  Thus, although the thermodynamics of the model in the far future is the same as that of our model of stable dS space, this is not true of its behavior at earlier times.
 
 The Jacobsonian THEFT of this model is Einstein gravity coupled to a perfect fluid stress tensor, which is the sum of a component with equation of state $p = \rho$ and a cosmological constant.   The metric is flat FRW, with scale factor $$ a(t) =[ \sinh (\frac{t}{3R})]^{\frac{1}{3}} .$$  It is a valid description of the thermodynamics of our model as long as the particle horizon radius is much larger than the Planck scale.  
 
 This model describes the history of our own universe in a broad brush fashion.  Note that flatness, homogeneity and isotropy are valid for generic initial states and we cannot be accused of fine tuning the Hamiltonian.  In fact, many will decry the fact that we do not have a unique Hamiltonian. We think that the answer to that complaint involves a very deep point, the necessity of locality to the conventional interpretation of quantum mechanics.
 In order to avoid interrupting the flow of our arguments, I will relegate that material to an appendix.  Suffice it to say that this model describes periods in the history of our universe, its beginning and end, in which there were no local excitations, and consequently (see the Appendix), only thermodynamic observables.   Many different mathematical models will give the same answer for the only real questions one can ask in such a universe.
 
 Finally, we consider a combination of the two previous models.  As far as the Hamiltonians of individual trajectories are concerned, the new model coincides with both of the old ones, up to time $N_{max}$. Now consider some particular point in the lattice of trajectories.   Within, and on the boundary of the tilted cube $N_{max}$ steps away from the central point, we stop the growth of the Hilbert space at $N = N_{max}$.  Outside, we let the Hilbert space grow indefinitely, as in the $p =\rho$ model.   The overlap rules are modified so that there is never any overlap between points in the interior of the tilted cube and points outside it.   The overlap between exterior points and the cube's boundary points never gets larger than the Hilbert space ${\cal H} (N_{max})$ of the boundary point.
 
 These overlaps rules are completely consistent.  The boundary is a spherically symmetric, marginally trapped surface.  Geodesics inside it are causally disconnected from those outside. The space-time geometry implied by this model is a spherically symmetric black hole, with dS interior,  embedded in the flat $p=\rho$ FRW geometry.  Geometries representing black holes embedded in FRW universes have a long and tortured history.  The McVittie solution\cite{mcvittie}  of the pure Einstein equations exists for every black hole horizon area and every FRW scale factor $a(t)$. It is a solution of the Einstein equations with some perfect fluid stress tensor. However, its horizon is actually a weak null singularity where second derivatives of the curvature diverge\cite{kaloperetal} rather than a smooth marginally trapped surface.  
 Nothing special appears to happen to this solution when $a(t) = t^{1/3}$, and it's not clear whether it is the appropriate geometry exterior to the trapped surface.  Our quantum model does have a rather abrupt transition, on the Planck scale, between the dS region and the $p = \rho$ region.
 We can take the $p=\rho$ McVittie solution, with the null singularity and its interior excised, and glue in a single horizon volume of the dS space with the same radius.  The resulting geometry satisfies the Israel junction conditions with a sensible though singular stress tensor on the interface.  I don't know whether it's necessary to find a non-singular solution, since our quantum model is defined with an abrupt transition at the Planck scale.
 
 We do not have quantum models, which might represent multiple black holes with different horizon sizes, each with its own interior dS space.  However, general properties of Einstein's equations assure us that if one such solution exists, then so do others, with widely separated black holes that have slow relative initial velocities.  Nothing would seem to constrain the sizes and other collective coordinates of such a solution, besides a few inequalities.  We will invoke such solutions in addressing certain meta-cosmological questions below.

\section{Holographic Cosmology of Our Universe}

The absence of localized excitations in all of the models of the previous section can be re-stated as an entropy problem.  All of those models maximize the entropy in a horizon volume at all times.  Our formulation of the theory of Minkowksi and eternal dS space realizes localized particles as low entropy states, satisfying of order $N$ constraints. This point of view is validated by the behavior of the cosmological horizon in the presence of localized excitations at the origin of the static patch- localized excitations {\it decrease} the entropy.  Thus, the question we want to ask is ``What kind of model will get the universe into a low entropy state, which has localized excitations of the eventual dS universe?"  .  This is a version of Penrose's famous question but has an unexpected twist to it.  Indeed, the classic discussions of inflation discuss its ``resolution of the homogeneity, isotropy and flatness {\it problems}" as if these were part of Penrose's question.  We have seen that they are not, and in fact that these properties follow for generic states of a large class of models, all of which can be thought of, in their early stages, as the most generic description of the beginning of a Big Bang universe.  Penrose's own formulation of the entropy problem in terms of the Weyl curvature stems, I believe, from the idea that black holes always maximize the entropy.  This is true locally in empty spaces, but a universe densely filled with black holes in fact has the relation $\rho \sim \sigma^2$ between energy and entropy densities and so its coarse grained description is the $p = \rho$ FRW model, which has zero Weyl curvature.  Thus, I believe the right way to ask Penrose's question about why the universe ``started" with low entropy is to ask {\it why there is something instead of everything}: why our final empty (but high entropy) dS space was preceded by a low entropy gas of localized excitations, rather than the high entropy but featureless $p = \rho$ state.

Given the building blocks introduced in the previous section, there is one obvious way to do this.  Our second model has many of the features attributed to ``inflation" in the QUEFT picture of that phenomenon. There are many horizon volumes of dS space, with independent, decoupled degrees of freedom.  Implicit in the definition of inflation is the assumption that eventually these independent regions all come into causal contact with any given time-like trajectory, and in particular that they can be detected by terrestrial detectors.  Our strategy then is to append the Hilbert spaces of $e^{3N_e}$ individual trajectories in our second model as tensor factors in the Hilbert space of a single one.  In principle, we want to do this for every trajectory in the part of space-time that is included in ``our universe" and impose the HST consistency conditions.  We have not yet been able to do this, and will concentrate on the Hamiltonian of a single trajectory.

We incorporate the boundary condition for an asymptotically empty dS space of radius $R$, with $N^2 L = \pi (RM_P)^2$, by insisting that after $N_e$ e-folds of inflation, with a horizon volume whose entropy is $n^2 L $, we have
$$ e^{3N_e} n^2 L = E^2 L < N^2 L  = 10^{123}.$$  The Hamiltonian we describe will evolve to that of a time-like geodesic in empty dS space, which has a preferred point, the origin of the static coordinates
$$ds^2 = - d\tau^2 (1 - r^2 / R^2) + \frac{dr}{1 - r^2 / R^2} + r^2 d\Omega_2^2 .$$In our lattice of trajectories, we choose this to be some particular point we call the origin of the lattice, and the horizon $r = R$ is a tilted cube of points 
$N\sqrt{L} $ steps from the origin.  The time coordinate $t$ of our Jacobsonian FRW model will approach $\tau$ in the infinite future, in a manner specified in \cite{DKS}.  The Hamiltonian, at each point of the lattice is that of the second model in the previous section, up to time $N_{max} = n$.  From that time forward we will concentrate on the Hamiltonian at the origin, and follow its evolution until the particle horizon has grown to the size $E$.  This is what is usually called ``the end of inflation".   Many inflationary horizon volumes are in causal contact.  I caution the reader that this terminology will be confusing for those used to the conventional picture of inflation, which is based on QUEFT.  In that picture, field theory degrees of freedom in MANY horizon volumes of dS space are assumed to be in causal contact.  This occurs because one considers modes which were originally of sub-Planck scale wavelength, in the original, un-inflated horizon volume.  This is inconsistent with HST.  Instead we will take many copies of our model of Big Bang-dS cosmology, and couple them together gradually in a manner described below.  We imagine those DOF as situated at the centers of tilted cubes of n steps, which tile a much larger tilted cube.  The space time geometry must be such that, at the time we call the end of inflation, the boundary of that large cube is a 2-sphere of radius $E^2 L$ . 

We will arrange those $E^2 L $ DOF on a ``fuzzy 3-sphere" with approximate $SO(1,4)$ invariance.  This is defined as follows.  A 3-sphere is a fiber bundle of two spheres over an interval $[0,\pi ]$.  At polar angle $\theta_k$ the two sphere has radius $ N_3 L_P\sin (\theta_k )$, and $N_3 L_P$  is the radius of the 3-sphere.  We define a fuzzy 3 sphere by discretizing the $\theta$ interval and using variables $\psi_i^A$ with appropriate values of $N_k$ to define the fuzzy two sphere at radius $N_k L_P =  L_P N_3 \sin (\theta_k)$.  The total number of DOF in our fuzzy 3-sphere is thus of order $$N_3^2 L \sum_0^P \sin^2 (\theta_k ) = E^2 L. $$ We choose the angles such that $N_3 \sin (\theta_k)$ is an integer multiple of $n$, which is possible since $E \gg n$ (the number of e-foldings is large).  In addition, $N_3 \sin (\theta_0 ) = n$ and the angles are distributed symmetrically around $\theta = \frac{\pi}{2}$.   

On each two sphere, we can introduce a basis of spinor harmonics consisting of maximally localized functions around a collection ${\bf \Omega_r}$ of $e^{3N_e}$ points.  The (fuzzy) localization scale is $1/n$.  That is, each point has fermionic pixel variables $\psi_i^A (P)$ that are $n \times n + 1$ matrices.  We choose the points to lie at the centers of the tiles of a geodesic grid on the two sphere with icosahedral symmetry.  The scale of the tiling is determined by the value of $\theta_k$, so that the two sphere at $\theta_k $  has $N_3^2 \sin^2 (\theta_k )$ points.   Clearly, for large $N_e$ these points become dense on the 3-sphere.  At each point, the variables are acted on by $SU(2) \times SU(2)$ acting separately on the rows and columns.  The diagonal $SU(2)$ is viewed as the stabilizer of the point in $SO(4)$, while the rest of the action can be used to parallel transport variables from point to point.   In this way, we can construct a Hamiltonian that formally transforms as the component of a $4-$vector that commutes with the diagonal subgroup of  $SO(4)$ in the large $N_e$ limit.
This Hamiltonian is viewed as the $J_{04}$ generator of $SO(1,4)$.  The rest of the generators are obtained from $J_{04}$ by conjugating with the emergent $SO(4)$ action.   There are many choices for $J_{04}$ that are $SO(3)$ singlets and transform as a component of a $4$ vector in the limit.  We further constrain them by insisting that in the limit $N_e \rightarrow\infty$,
$$[J_{0M}, J_{0N} ] = i J_{MN}.$$  That is, we insist that in the large $N_e$ limit, our system has a unitary representation of $SO(1,4)$.  In, this limit, our variables live on a fuzzy sphere whose radius is going to infinity and it is entirely plausible that the conformal group of the sphere be representable in terms of them.  We do not know if there are multiple ways to do this, or whether the resulting limit has the $Z_2$ TCP symmetry of global dS space as well.   We expect the corrections to $SO(1,4)$ symmetry to be of order the spacing of our grid on the 3-sphere, which is $o(e^{-N_e} )$.  

In our analysis for CMB fluctuations, we will use only the approximate symmetry properties of the model.  In particular, the scalar and tensor fluctuations of cosmological perturbation theory are assumed to arise from expectation values of commuting operators on the 3-sphere in some $SO(1,4)$ invariant density matrix.  In our model, for large $N_e$, the density matrix evolves  as
$$\rho \rightarrow e^{-i J_{04} t} \rho e^{ i J_{04} t} ,$$ for a long time.  $\rho$ can be written as a sum of operators transforming in different representations of $SO(1,4)$.  Furthermore, its initial conditions are approximately $SO(4)$ invariant because the individual systems of entropy $n^2$ are all prepared in the same initial state, by evolution prior to the time these systems are coupled together\footnote{This synchronization is a consequence of the consistency conditions of the HST model prior to the time that these systems come into the horizon of the central observer.}.  Time averaging the  density matrix will therefore project out the singlet component of $\rho$.
 
The fact that the fluctuation operators commute follows from our hypothesis that the variables at different points on the grid are constructed from bilinears of independent fermionic variables.   

\section{Meta-Cosmological Problems}

The model of the previous section has three parameters, $n, E, $ and $N$, satisfying the {\it a priori} constraints
$$1 \ll n \ll E < N .$$  It gives rise to a universe, which evolves from a $p=\rho$ Big Bang, through a period of inflation characterized by $n$ and $E$, to a dS state characterized by $N$. It generates approximately scale invariant adiabatic fluctuations, whose primordial magnitude is $\frac{1}{\epsilon n}$.   There appears to be no mathematical barrier to constructing a model of this type for every value of these parameters consistent with the above inequalities.   How did Nature choose the values that fit the data of our universe?

The HST answer to this question is {\it environmental selection}.  Recall the model of an asymptotically dS cosmology as a black hole whose interior contains the cosmology, embedded in a $p = \rho$ background.  If one such black hole exists general properties of Einstein's equations guarantee the existence of multi-black hole solutions, each of whose interiors is a different cosmology of the class constructed in the previous section.  The parameters of those models then become subject to environmental selection.  Only those black holes whose internal cosmology can support complex life, will be observed by complex life forms.  

In addition to the internal structure of black holes, multi-black hole solutions will have collective coordinates corresponding to the initial relative positions and velocities of the holes, in the background FRW geometry.  Thus they can suffer collisions.   No observation within a single black hole interior, can determine the parameters governing the rate of collisions of that black hole with others, until the collision occurs.   We will use this freedom to eliminate the philosophical objections to eternal dS cosmologies, which go under the rubric ``Boltzmann's brains". 

First to the determination of cosmological parameters:  The eternal dS model with radius $N$ approaches, as $N\rightarrow\infty$ a four dimensional super-Poincare invariant model, with minimal SUSY and an R symmetry, which has no continuous moduli.  There is no such model found in the entire catalog of string compactifications and zero c.c. limits of AdS/CFT .   In addition, if we perform exercises like finding $R$ symmetric stationary points of the Gukov-Vafa-Witten\cite{GVW} superpotential, we find that R symmetric points are rare.   

Recall also that in HST, moduli are discrete, with continuous moduli arising only in limits where there are length scales much larger than the higher dimensional Planck scale.  Generic string theory models, away from such extreme regions of moduli space, can be thought of as living in $11$ dimensions. The ratio between the Planck scale and the unification scale is related to the volume of the extra dimensions by $$10^4 = (\frac{m_P}{m_U})^2 = (R_D m_D)^7 ,$$ where $R_D m_D$ is a typical internal dimension in Planck units.  This gives $R_D m_D \sim 3.73$, which is not terribly large.  We do not expect approximately continuous changes of moduli in the vicinity of the model of the real world. Finally, note that ratios of low energy scales to the Planck scale are explained by asymptotic freedom, which implies exponential sensitivity to the values of high energy couplings.
 I take this as evidence that if there is an HST model consistent with the phenomenology of the standard model of particle physics, then any other super-Poincare invariant limit of a theory with finite c.c., will have low energy physics wildly different from that of the standard model. 
 
 The connection $m_{3/2} \sim \Lambda^{1/4} $, implies that the scale of SUSY breaking in an HST model is low, which means that in order to account for experimentally acceptable gaugino masses in QUEFT, SUSY breaking must be mediated directly from a new strongly coupled sector, containing particles charged under the standard model.  The masses of these particles and the scale of the strongly coupled sector are determined by the scale of SUSY breaking.  Thus, dialing the c.c. makes drastic changes in the low energy physics of particles, atoms, molecules and stars.
 It is plausible that $N$ is fixed by the requirement that standard model physics gives rise to the standard elements and properties of stars.  In the Pyramid Schemes\cite{pyramid}, the only TeV scale models I've found that are consistent with the properties of HST and phenomenology, this is certainly the case. It is also likely, in the context of a highly constrained system, with no moduli, that these requirements also fix the physics of dark matter, and its abundance. The current model of dark matter in the Pyramid Schemes, relies on a primordial asymmetry in baryonic excitations of the new strong gauge group, which is tied to the ordinary baryon asymmetry of the universe.
 
 If that is the case, then Weinberg's galaxy formation bound\cite{wbl} becomes 
 a lower bound on the size of primordial fluctuations of the form
 $$\frac{1}{\epsilon n} > (\frac{729 \Lambda}{500\rho_0})^{1/3}.$$ Given the known values of $\Lambda$ and $\rho_0$, which by the argument of the above paragraph we assume constrained by the HST model of particle physics and environmental selection of atoms, nuclei and stars, and the {\it a priori} constraints $n \gg 1$, this gives a rather good determination of the amplitude of primordial fluctuations.  
 We note also that this argument assumes $E \ll N$.  We have not yet developed an HST model of the radiation and matter dominated eras of cosmology, but its predictions will surely be very close to the standard cosmological model, if it is not to be ruled out.  Weinberg's bound assumes that those eras indeed followed the standard model.  
 
  We have to check that the latter inequality is consistent with the fundamental requirement that the causally connected inflationary fluctuations predicted by the model, have enough information in them to be identified with the fluctuations in the CMB.  The entropy in fluctuations is roughly
 $$ \delta T T^2 R^3 = 10^{-5} T^3 R^3 \sim 10^{84},$$ where $T$ is the CMB temperature and $R$ the radius of the cosmological horizon.  This is the lower bound on $E^2 L$, the total entropy at the end of inflation.  Recalling that $$E^2 = e^{3N_e} n^2 $$ and $$\epsilon n \sqrt{L} = 10^5 ,$$ this gives $$e^{3N_e} 10^{10} > \epsilon^2 10^{84} .$$ Thus $$N_e > \frac{74 {\rm ln}\ 10 + 2 {\rm ln}\ \epsilon}{3} = 55.26.$$  In the final equality, we've taken $\epsilon = .1$, which is what the data indicates roughly, in the context of the HST model.
 
 This gives a value of $E$ comfortably below $N$.  I suspect that further investigation will show that $E$ has to be close to this lower bound, in order for galaxy formation to proceed in a conventional manner.
 
 Finally, we can use the assumption that the multi-black hole solutions in the $p=\rho$ universe correspond to real quantum HST models, to resolve a philosophical problem that has been posed for eternal dS spaces.  While we've been at pains to construct a model of the universe in which the conventional low entropy initial state of small local fluctuations on an FRW background arises without tuning, there are those who would criticize the model because after the time when our local group of galaxies collapses into a slowly evaporating black hole, there are an infinite number of local thermal fluctuations that occur, and produce short lived observers, with our memories but abruptly truncated future histories.   Indeed, there are an infinite number of points in the history of eternal dS space in which much of our observable universe will fluctuate into existence and remain in existence for a long time
 during which observers will verify that the universe is not quite what they would have expected on the basis of the standard cosmological model.   This is called, by some, the problem of Boltzmann Brains.
 
 To get a rough estimate of the probability of one of these exotic processes, assume the smallest mass that such a problematic observer could have is $m$.  Then the probability per unit time per unit volume of the fluctuation is less than
 $$ e^{- 2\pi m R} e^{- 4\pi m^2 / M_P^2} t^{-4} .$$ The first exponential is the thermal probability of a thermal fluctuation of mass $m$, while the second is the probability that the fluctuation is not a black hole.  For $m$ of order the Planck mass, which is about $10^{-5} $ grams, this is $ \sim e^{ - 10^{62}} t^{-4} $.  Letting $t$ vary between the Planck time and the current age of the universe changes the number in the exponent by a negligible additive constant of order $10^2$.  In other words, the time scale for worrying about this ``problem" is so much longer than anything in observable physics that one can easily modify the theory in an infinite number of possible ways, which will eliminate the problem without changing any observable predictions of the theory.
 
 This is true even as a matter of principle.  Since most of the degrees of freedom in dS space are not localizable, and only localizable DOF can be used to construct measuring apparatus, there is a bound on the time scale over which the collective coordinates (pointers) of any conceivable apparatus will suffer quantum fluctuations and become useless for recording data.  In the language we used in the appendix, after this time scale the entire preceding history of the universe will have ``un-happened".  A subsystem of the universe weighing 
 $\sim 10^{30} M_P$ or more can only fluctuate into existence on a time scale longer than this un-happening time.
 
 A particular example of how we can avoid Boltzmann's Brains is afforded by an infinite number of possible multi-black hole models.   The model must contain enough black holes that there is a non-zero probability of having values of $n,L,N,E$ compatible with those in our model of the universe.  The relative initial positions and velocities of these black holes in the $p =\rho$ background must be such that at least one universe with those values of the parameters survives through a time long compared to the current age of the universe.   Finally, collisions must be such that a collision between ``our" black hole and another occurs before many BB's are nucleated.  The extremely long time for the nucleation process makes it easy to satisfy these two conditions.
 
 I do not have a theory of precisely what happens in such a collision, but in the best case scenario, the combined system will settle down to dS space with a much smaller value of the c.c. , since its entropy has increased.  Even if exactly the same local micro-physics is valid, the value of the SUSY breaking scale will be much smaller, while other scales are unchanged.  This will lead to very different atomic and nuclear physics, and it's very unlikely that organisms like ourselves will exist.  To be more precise one needs an effective model of low energy cosmological SUSY breaking, like the Pyramid Scheme\cite{pyramid}.   In that model it is clear that a significant lowering of the c.c. completely wrecks nuclear physics and chemistry.  
 
 Just as significantly, lowering the c.c. by a factor $A$ reduces the nucleation probability from $e^{- K} $ to $e^{ - \sqrt{A} K}$ .  So even if we assumed low energy physics and chemistry was untouched, a black hole collision rate significantly smaller than the current inverse age of the universe but significantly larger than the BB nucleation rate in a hypothetical eternal dS space with our value of the c.c., would in fact predict that BB nucleation never occurred.  
 
 To summarize then, we have modeled the very early universe up to the age of inflation as an almost featureless fluid, characterized by a small number of parameters, $n,N,E,L$.  On the large scale it looks like an asymptotically dS universe, which undergoes a period of inflation to generate localized fluctuations.  The whole system is embedded inside a marginally trapped null surface in the flat $p=\rho$ FRW.  Einstein's equations have solutions with multiple black holes of this type, and we have conjectured that there are quantum HST models corresponding to those.  This allows us to use environmental selection criteria to choose the underlying parameters (an important part of that selection is based on the connection between the value of $N$ and the scale of SUSY breaking).  It also allows us to construct an infinite number of models which have no Boltzmann brain problem.  However, there is no observation we can make, which verifies whether or not the model actually does solve that problem, because the time scale on which the solution operates is much longer than the age of the universe.

\section{Cosmological Observables and $SO(1,4)$ Invariance}

We have argued above that when the number of e-foldings is large, the density matrix of our HST model has an approximate conformal invariance, which means that all predictions can be written in terms of correlation functions of operators that transform in unitary representations of $SO(1,4)$.  Violations of the symmetry in these correlation functions should be small, of order $$e^{-N_e}$$ for correlation functions involving a small number of operators.  In this limit we have also localized our operators on a three sphere.  Finally, the operators all commute with each other, since they represent fluctuations of independent systems, each of which has entropy $n^2 L$.  The size of the fluctuations is $(n^2 L)^{- k/2}$ for $k$ point functions with low $k$.   

The hydro-dynamics of the model can be described by a Jacobsonian THEFT, which we view as a small perturbation of a background FRW geometry, with scale factor $a(t)$ .  $H(t)$, the background Hubble parameter, should undergo $N_e$ e-folds of inflation, followed by sub-luminal expansion, which brings the entropy of the cosmological particle horizon to $E\sqrt{L}$, according to the Bekenstein-Hawking formula.  The slow roll parameter, $\epsilon = \frac{\dot H}{H^2}$ is always $\ll 1$.  
The inhomogeneous fluctuations are stochastic fields, whose low order statistics reproduces that of the quantum operators $S$ and $T$.  The exact $SO(3)$ rotation symmetry of the underlying model implies that the vorticity of the cosmological fluid vanishes.  In this case \cite{lythliddle} one can always choose a co-moving gauge for the fluid, and the fluctuations are part of the metric field.  The metric takes the form

$$ds^2 = -dt^2 N^2 + h_{ij} (dx^i + N^i dt)(dx^j + N^j dt) ,$$ where the lapse and shift are determined by solving the constraint equations.  In the co-moving gauge
$$h_{ij} = a^2 (t) [(1 + 2\zeta )\delta_{ij} + \gamma_{ij}],$$ where $\gamma$ is transverse and traceless.  Note that the zero momentum component of $\zeta$ is just an infinitesimal rescaling of the background.  This leads to Maldacena's squeezed limit theorem\cite{malda}, which relates 3-point functions with a zero momentum $\zeta$ to the scaling behavior of two point functions.  

The scalar fluctuation $\zeta ({\bf x} , t) $ is just the proper time difference between points with the same coordinate label, on infinitesimally close co-moving hypersurfaces.  We can write
$$ \zeta = H \delta\tau = \frac {H \delta H}{\dot{H}} = \frac{H^2}{\dot H} \frac{\delta H}{H} . $$ The first factor is the inverse of the slow roll factor $\epsilon$, while the second is the relative fluctuation in the space-time curvature.
$\gamma_{ij}$ is also a curvature fluctuation.

As in conventional hydrodynamics, the fluctuations of these hydrodynamic variables $\gamma$ and  $\delta H / H$, are related to expectation values of operators in the underlying quantum theory.  In HST, we've argued that these should be commuting operator functions on the 3-sphere, transforming in unitary representations of the dS group $SO(1,4)$. The $k$ point functions of these operators should be of order $ (n\sqrt{L})^{-k}$, since they originate as fluctuations of a local thermalized system with entropy $n^2 L$. 

The combination of $SO(1,4)$ invariance and the squeezed limit theorems, determines all 3 point functions involving $\zeta$ in terms of the scaling behavior of two point functions.  The small size of measured scaling violations in the scalar two point function then implies both that the $\zeta - \zeta$ correlator is much larger than the $\gamma \gamma$ two point function (recall the explicit factor of $\epsilon$ relating $\zeta$ to $\delta H/ H$), and that all 3-point functions involving $\zeta$ are too small to be measured by current experiments.  

A lot of recent work has shown that a large class of conventional slow roll inflation models will satisfy a set of rules closely analogous to those of the HST models\cite{maldaetal}.   Although there is some confusion in these papers about the precise meaning of ``conformal invariance" and the relation to conventional conformal field theory, their results differ from those of HST in only the following observable ways:

\begin{itemize}

\item There is an additional breaking of $SO(1,4)$ invariance, which comes from the evaluation of the quantum fluctuations in terms of QUEFT in a non-dS background geometry $H(t)$.  The magnitude of the constants in these two point functions can be computed explicitly, while in HST they can only be estimated.

\item The scalar fluctuation operator in QUEFT has dimension corresponding to a bulk scalar field of mass zero.  It lies at the extreme end of the complementary series of unitary representations of $SO(1,4)$.  In HST we can choose {\it any} operator in the complementary series.  The principle series representations are not allowed.  The corresponding operators are not Hermitian and there would be two complex conjugate pairs of three point functions, whose physical meaning is obscure.  Despite these differences, the freedom to choose different models for $H(t)$ in QUEFT and HST means that we can fit the scalar two and three point functions in identical ways in both models.

\item  The tensor two point function in HST has no tilt.  While this is an unambiguous statement, the tilt is predicted to be $r/8$ in slow roll models and $r$ is known to be less than about $.1$.  Furthermore, we have not yet detected the tensor two point function.  Both HST and QUEFT models predict it to be smaller than the scalar two point function by a factor of $\epsilon^2$.  In HST it is predicted to be at the edge of visibility by the Planck experiment, while in QUEFT it could be this large, or many orders of magnitude smaller.  

\item In QUEFT, only one of the three allowed $SO(1,4)$ covariant forms for the tensor three point function appears in leading order in the $H/m_P$ expansion.  The second parity conserving term appears at the next order, but should be smaller than the first if the QUEFT expansion is valid.  The parity violating term cannot appear in any order of the expansion because the density matrix in QUEFT is diagonal in the same basis as the operators whose expectation value is being computed.  There is no such restriction in HST, and we would expect all covariant forms to be present with comparable weight, unless parity is imposed on the model as a fundamental symmetry.  
Unfortunately, this rather unambiguous prediction is probably below the threshold of current experiments.  All the tensor 3 point functions are smaller than the as yet undetected tensor two point function by a factor of $(H/m_P) \sim 10^{-4} $ or less.

\end{itemize}

The conclusion of this analysis is a bit disheartening.  All of what is known experimentally about the CMB and large scale structure can be explained in a rather general framework of cosmological perturbation theory and approximate $SO(1,4)$ invariance.  Many inflation models {\it do not} fit into this framework, and those can be falsified, but at the moment it appears impossible to distinguish between a large class of models of single field slow roll inflation, and the HST models, whose conceptual basis is radically different.  Careful measurements of tensor two point functions, and less careful measurements of tensor 3- point functions, would do much to improve this situation.

\section{Conclusions} 

Holographic Space-Time provides a formalism for quantum gravity which is as local as the holographic principle allows it to be.
In a given causal diamond, whose area is $N^{d-2}$ in Planck units, only states satisfying of order $N^{d-3}$ constraints, behave approximately like the particles of quantum field theory.  The generic state has energy of order $N^{- (d - 3)}$ and is a non-localized excitation on the horizon.  When these constraints are not satisfied, in a causal diamond whose proper time is much less than the age of the full space-time, external observers see the excited diamond as a black hole.
QUEFT cannot account for the detailed quantum properties of even large smooth black holes, because it leaves out the horizon DOF. However, Jacobson's results assure us that space-time geometry {\it does} encode the hydrodynamic properties of such systems correctly.  

The HST model of the early universe is also that of a maximal entropy state which is not well described by QUEFT, even when the energy density is low compared to the Planck scale.  An inflationary era introduces localized fluctuations, and explains ``why there is something (local) instead of everything (unconstrained horizon DOF)".   The model produces CMB fluctuation spectra, which are at present indistinguishable from the (conceptually very different) single field slow roll inflation models. Unfortunately, this is partly because all of the current data can be explained in a general framework, without reference to particular models\cite{holoinflation}.
The description of the particle physics dominated era of the universe that follows inflation has not yet been accomplished within the HST formalism.  I suspect that it will, for the most part, give results indistinguishable from standard cosmology, but the issue of cosmological phase transitions may hold some surprises. 

By far the most interesting aspect of HST, from a phenomenological point of view, is its prediction of a very low scale of SUSY breaking, embodied in the formula for the gravitino mass $$m_{3/2} = K (\frac{M_P^2\Lambda}{M_U^2})^{1/4}, $$ with $K$ of order $1$.   When combined with gauge coupling unification and the failure to observe particles beyond the spectrum of the standard model, this puts very strong constraints on QUEFT models of Tera-scale physics.  I hope that the next run of the LHC will provide some evidence for the surviving models, rather than ruling them out entirely.

\section{Appendix - Quantum Mechanics and Locality}

The theory of decoherence provides a sensible interpretational framework for the mathematics of the quantum theory.   My understanding of it goes as follows.  Consider a Hamiltonian with quasi-local interactions between a set of localized variables.  To be definite, let it be a theory defined on a regular lattice in $\geq 3$ dimensions\footnote{This restriction on the dimension is made to avoid discussing special infrared properties of one and two dimensional systems, but is probably unnecessary.}  The interactions do not have to be nearest neighbor but should fall off rapidly with lattice distance.

In many cases we can prove, but it is reasonable to conjecture in general, that the density of states grows exponentially with a power (less than one and greater than zero) of the energy.   Above a certain gap, the entropy in a fixed small energy band scales like the volume, and generically the dynamics is ergodic in some sense of that word.  That is to say, averages of products of local operators, in the density matrix obtained by averaging the Schrodinger state over a microscopic relaxation time, are well approximated by their averages in the microcanonical density matrix for that band.  

In such systems, averages of local operators over large volumes, $V_i$ of the lattice have the properties of {\it collective coordinates}:

\begin{itemize}

\item Their dynamics is slower than that of the microscopic degrees of freedom by a factor of $1/V_i$.

\item Their uncertainties are small.  The commutator of two such averages is, because of locality, only of order $V_i$, while their anti-commutator is of order $V_i V_j$ .

\item Most importantly, if we've chosen an energy band sufficiently far above the gap, then the number of states for which the collective coordinate has approximately the same expectation value is $o(e^{V_i})$.  Furthermore, the dynamics of the system is such that one is making rapid transitions between the states of this ensemble on a microscopic time scale.  As a consequence, quantum interference terms in the calculation of expectation values is {\it super-exponentially } small.  This is because, for what we call a macroscopic region, $V_i > 10^{20}$.

\end{itemize}

The quantum dynamics of collective coordinates thus resembles classical dynamics with $o(\frac{1}{\sqrt{V_i}} )$ statistical fluctuations, which obey classical statistics, and in particular, satisfy Bayes' rule, with super-exponential accuracy.   Given Bayes' rule, we can do Bayesian conditioning of the probabilities for these collective variables to be in one of their allowed states.  This is what is called {\it collapse of the wave function}.  

To say this all in a phrase borrowed from my colleague Scott Thomas: {\it In quantum mechanics, Objective Reality is an emergent phenomenon.}  The essence of QM is that probability is forced on us, because not all the variables, which appear in the equations of motion, can have definite values in the same state.  This gives a mathematically defined probability theory in which the probability for any quantity can be calculated in any state of the system. The state of the system satisfies a deterministic equation of motion, and its initial conditions are determined in terms of the initial values of a set of quantities which {\it can} all be determined at the same time.

Baye's rule is the key feature of a probability theory, which lets us imagine that there is an objective reality that evolves deterministically, but that we have simply not included all the relevant data in our initial conditions.  It is the rule that allows us to say ``probability theory tells us about measurement uncertainty, but once we've actually done a measurement, we know which of the many possible outcomes the theory predicted {\it could have happened}, {\it actually happened}". It is not, in general, satisfied by QM probabilities, but it is satisfied, with super-exponential accuracy, for the quantum probabilities predicted for collective coordinates.

The emergent nature of ``objective reality" in QM, is vividly illustrated by the phenomenon of {\it unhappening}.  Suppose you write your name on a piece of paper, and then explode a bomb, which destroys the paper and yourself, and any other nearby object that has collective coordinates.  The information about that classical event, including what your name was, whether you wrote it, what color the ink was, {\it etc.} is encoded in the quantum wave function of the debris from the explosion, much of which consists of elementary particles that have no collective coordinates.   The objective reality of your having written your name is now scattered non-locally in an ever growing sphere of photons that bounced off the paper before you set off the bomb.
Even if we could set up a conspiracy among many local observers, to measure properties of the individual photons, no single measurement would actually determine the multi-photon state, because that state is not an eigenstate of the quantities being measured.  The statement ``Tom wrote his name on a piece of paper in purple ink ten million years ago", is no longer true or untrue.   It is one of myriad statistical possibilities that are compatible with the limited information about the multi-photon state, which can be gleaned in any collection of distributed local measurements.  The event of name-writing has {\it un-happened}.  There is no actual measurement, which can measure the projection operator on the subspace of states in which the event definitely happened.

As we have discussed in the text, in HST, there is a limited amount of information that can be encoded in states, where the kind of lattice model described above is a good approximation.  More generic states of the system have the properties of black holes:  their only collective coordinates are truly thermodynamic averages over the whole system.  

This makes the question of low entropy of the early universe a crucial one for the emergence of any kind of classical physics, but also points to an environmental selection answer to the question of {\it why} the entropy was low.  Certainly, no kind of information rich environment can occur, without this low entropy starting point.  The time dependent nature of the Hamiltonian in HST, allows us to choose models in which at an intermediate time, the low entropy state, in which the the state of the universe remains a pure tensor product state in the same basis for an extended period, is forced on us by the dynamics.

\section{Acknowledgements}

I would like to thank W. Fischler for his collaboration on much of the work discussed in these lectures. I'd also like to thank the organizers of the Davis, Nottingham, and Benasque conferences, and the participants of the Benasque workshop, for affording me the opportunity to talk about this material.
 The work of TB was supported by the DOE.

\end{document}